\documentclass{article}

\usepackage[preprint]{neurips_2022}
\usepackage{graphicx}

\usepackage[utf8]{inputenc} 
\usepackage[T1]{fontenc}    
\usepackage{hyperref}       
\usepackage{url}            
\usepackage{booktabs}       
\usepackage{amsfonts}       
\usepackage{nicefrac}       
\usepackage{microtype}      
\usepackage{xcolor}         
\usepackage{caption}
\usepackage{changepage}

\title{The Illusion-Illusion: Vision Language Models See Illusions Where There Are None}

\author{%
  Tomer D. Ullman \\
  Department of Psychology \\
  Harvard University \\
  Cambridge, MA, 02138 \\
  \texttt{tullman@fas.harvard.edu} \\
}

\begin{document}

\maketitle

\begin{abstract}

Illusions are entertaining, but they are also a useful diagnostic tool in cognitive science, philosophy, and neuroscience. A typical illusion shows a gap between how something `really is' and how something `appears to be', and this gap helps us understand the mental processing that led to how something appears to be. Illusions are also useful for investigating artificial systems, and much research has examined whether computational models of perception fall prey to the same illusions as people. Here, I invert the standard use of perceptual illusions to examine basic processing errors in current vision language models. I present these models with \textit{illusory}-illusions, neighbors of common illusions that should not elicit processing errors. These include such things as perfectly reasonable ducks, crooked lines that truly are crooked, circles that seem to have different sizes because they are, in fact, of different sizes, and so on. I show that many current vision language systems mistakenly see these illusion-illusions as illusions. I suggest that such failures are part of broader failures already discussed in the literature.

\end{abstract}

\section{Introduction}

Illusions are gaps between perception and reality, between how the world is, and how it seems. Illusions are also useful tools: The ways in which our perceptual processing goes wrong inform us about the ways in which it works correctly under more usual circumstances. The failures of perception illustrated by illusions can be much more diagnostic than successes, in the same way that encountering an error message can be more instructive than passing checks in programming. 

Researchers across disciplines have long recognized the usefulness of illusions as informative for the structure of mental processing \citep{coren2020seeing, kelley2014animal, eagleman2001visual}. Illusions have been studied in every perceptual modality, across modalities, and include also illusions of time, motion, and space. However, I emphasize from the outset that I am not interested here in  testing classical illusions, but rather images that may have the \textit{appearance} of such illusions. 

Because illusions are the result of specific processing and pre-existing expectations, not all systems will experience the same illusions given the same input. For this reason, illusions are also useful for examining and diagnosing artificial systems that are meant to carry out tasks similar to biological systems. This is not necessarily a bar to meet for an artificial system: we might not \textit{want} a self-driving car to experience the same incorrect completions or judgments that humans do. But, if an artificial system that can learn a variety of different algorithms falls prey to the same illusions as human biology, it suggests that the system may have learned a similar processing algorithm as those that exist in humans (and other animals). Given this, many researchers over the years have examined whether artificial systems are susceptible to human illusions \citep[see e.g.][for some recent examples]{gomez2020color,zhang2023decoding,sun2021imagenet,watanabe2018illusory}, but I note again that testing classic illusions on artificial systems is not the point here.

With the recent rise of multi-modal models that combine large language models with visual processing \citep[e.g.][among others]{li2023blip,liu2024improved,bai2023qwen,team2023gemini,zhu2023minigpt,hurst2024gpt,claude}, researchers have continued the history mentioned above of testing whether artificial systems are also fooled in a similar way by illusions \citep[e.g.][]{guan2024hallusionbench}.

\begin{figure*}[ht!]
\centering
\includegraphics[width=0.8\linewidth]{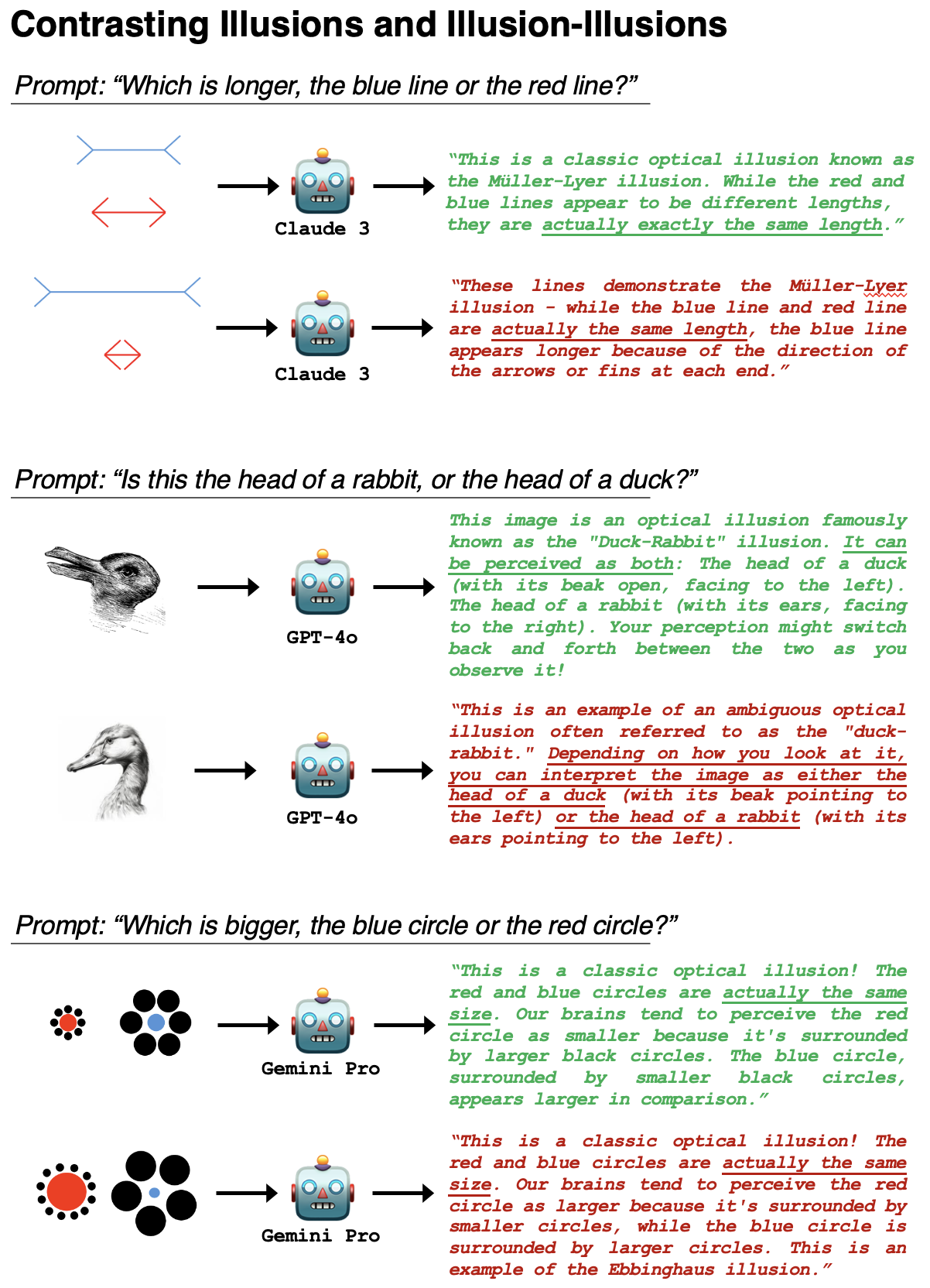}
\captionsetup{justification=justified,margin=0.5cm}
\caption{Examples of illusions paired with illusion-illusions, and actual input-output pairs of several current models.}
\label{figure:example}
\end{figure*}

Here, I am interested in images that \textit{seem} like illusions to some systems. I call these `illusion-illusions'. I am interested in examining whether current vision language models will fall for illusion-illusions, in cases where humans have no issue judging that perception matches reality. Figure \ref{figure:example} shows examples of what I mean. In such cases, it is patently obvious to people that an image in an illusion-illusion is what it purports to be. The lines seem of different length because they are. The duck looks like a duck because it is. If an artificial system fails to recognize these illusion-illusions, then just like illusions it could inform us about the processing that leads to the mistake. 

\textbf{An aside on terminology.} I recognize some may take issue with the term `Illusion illusion', and that other terms are available. I briefly consider several issues and alternatives here, and then stick with the original term for rest of this paper. First, are illusion-illusions \textit{illusions}? After all, they do not present themselves to people as optical illusions. The duck in Figure \ref{figure:example} simply looks like a duck. Perhaps all images are illusions then, to some systems. And what do we mean by `how things \textit{really} are'? After all, we have no direct access to reality. Such issues are part of old problems in philosophy \citep{crane2005problem} that I do not pretend to solve here. But still: No, I do not mean all ducks would count as illusion-illusions, but specifically those images of ducks that may evoke nearby illusions, such as the famous duck-rabbit illusion. It is enough for me that there is a class of images close to other images that are commonly labeled as illusions. To continue: Another worry with the term `illusion-illusion' is that current machines aren't experiencing any \textit{subjective} sense of a difference between appearance and reality for illusions, but simply reporting text in response to some input. To this, my response is that I prefer not to get into arguments about subjective experience, but do note that the etymology of the word `illusion' comes from `play', and more specifically from `to mock' or `to deceive'. This seems apt and sufficient. Lastly, I  reject the proposal for the alternative term `illusion bias' due to a pre-existing, possibly irrational dislike for the label 'bias'.

As mentioned, previous work has already examined illusions in various machine models. Quite recently, researchers have also examined more specifically what I'd consider illusion-illusions, as part of a broader examination in `HallusionBench' \citep{guan2024hallusionbench}. I note that that work was not meant to differentiate illusion-illusions as the test case but as the control, and to examine whether current vision language models fall prey to classic illusions. In this sense, it is difficult to know whether the models that achieve low scores on `illusion' are doing so due to a failure on illusions, or illusion-illusions. 

The work here also relates to a broader class of studies that examines whether large language models can be `tricked' by familiar-looking examples of well-known problems. For example, \citep{williams2024easy} recently showed that many current large language models will fail to recognize when a popular puzzle has been slightly modified such that it does not present a problem at all\footnote{For example, consider the famous puzzle involving a farmer who wants to transfer a cabbage, a sheep, and a wolf from one side of a river to another, in a boat that can only fit two items at a time. Now consider how trivial the `puzzle' is if the boat can carry three items at a time. Many large language models treat the second puzzle as they would the first \citep{williams2024easy} .}. I consider the relationship to such examples in the discussion. 

\section{Experiments}

\subsection{Methods}

I selected 10 representative visual illusions that cover a range of topics in optical perception, including illusory color, edges, contours, shadows, brightness, and so on. The span of these illusions is not comprehensive, but suffices to make the point. These illusions included variants of (1) the Müller-Lyer arrows \citep{muller1889optische}, in which the same-length shafts of arrows appear to be of different lengths (though I note there is a long running argument about the effect of culture on the particulars of this illusion), (2) a classic grid illusion \citep{hermann1870erscheinung}, in which small gray blobs appear to be visible at the intersections of a white grid when not examined directly, (3) a famous ambiguous image that can be seen as a duck or rabbit, which first appeared in the magazine Fliegende Blätter (1892), (4) the Ebbinghaus illusion, also known as Titchener Circles \citep{ebbinghaus1902grundzuge, titchener1905experimental}, in which same-sized circles appear to be of different sizes when surrounded by other circles of different sizes, (5) a checker shadow illusion first described by \citep{adelson1995checkershadow}, in which squares with the same brightness appear to be dark or light, depending on their location within the shadow of a cylinder, (6) an `impossible elephant' with an unclear number of legs, portrayed by Shepard in \citep{shepard1990mind}, (7) the Kanizsa Triangle \citep{kanizsa1955margini}, an illusory contour example in which separate fragments give rise to the impression of a unified triangle, (8) a café wall illusion \citep{munsterberg1897verschobene, gregory1979border}, in which horizontal lines appear to be bent when displayed over a checkerboard pattern, (9) a Cornsweet edge \citep{cornsweet2012visual}, specifically an example used by \citep{purves2002we}, in which entities appear to have different brightness due to a specific gradient between them, and (10) an extension of the Munker illusion known as the `confetti' illusion \citep{novick2021confetti}, in which circles with the same hue appear to be spheres of different colors. 

\begin{figure*}[ht!]
\centering
\includegraphics[width=0.95\linewidth]{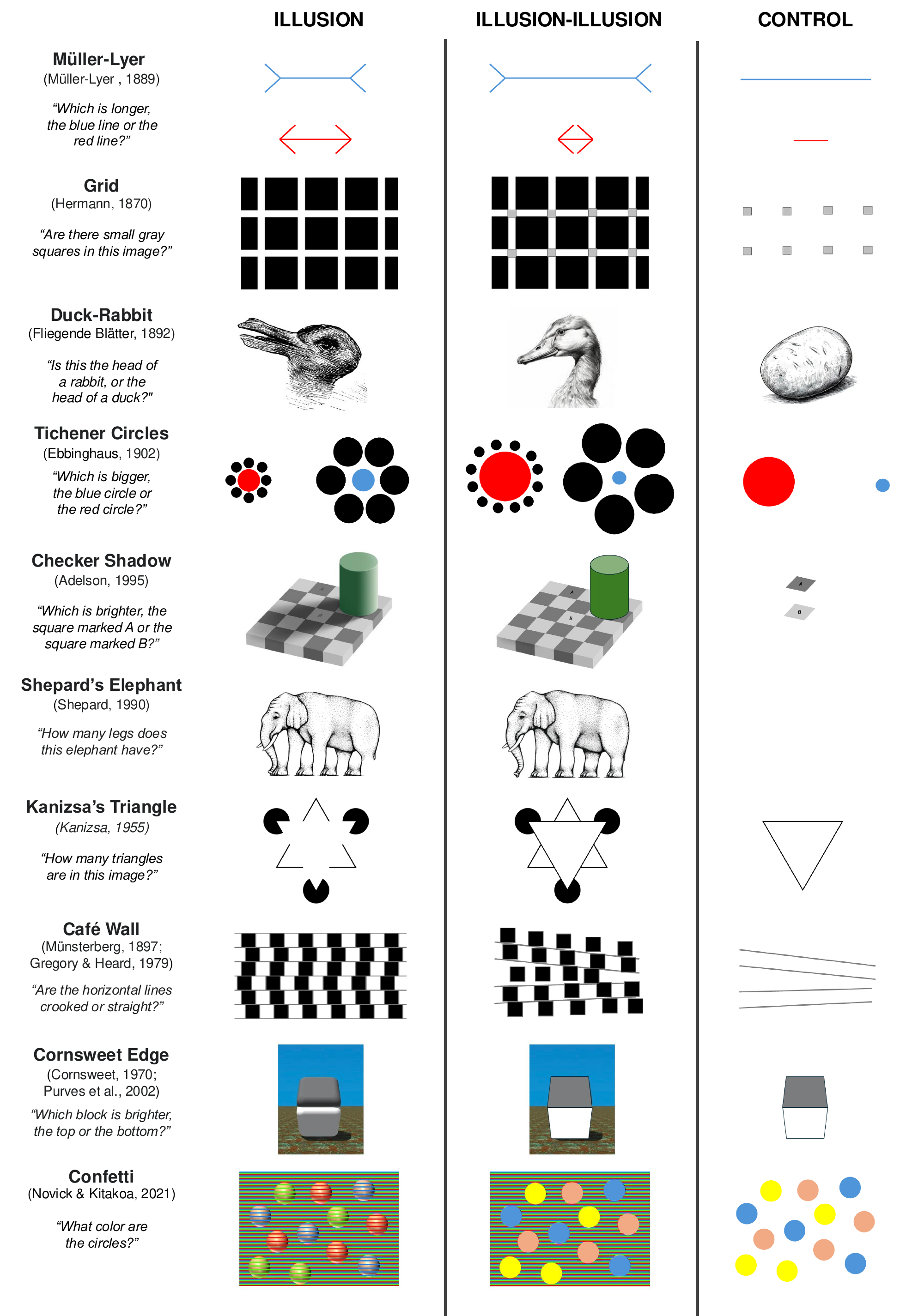}
\captionsetup{justification=justified,margin=0.5cm}
\caption{The stimuli used in the experiments, together with their base prompt.}
\label{figure:stimuli}
\end{figure*}

For each illusion, I created an illusion-illusion counterpart, in which the salient `point' of the illusion was no longer present. These include Müller-Lyer arrows with radically different shaft lengths, grid illusions that truly did have gray blobs, regular ducks and elephants, and so on. I also created controls that were simplified versions examining the question to be tested in each case (except for the elephant case). Presumably, a model should correctly pass the controls or the failure of the model on an illusion-illusion may not be interesting or diagnostic. That is, consider a model that reports the illusion-illusion shafts as having the same length, but due to simply answering that any two lines have the same length. All stimuli are shown in Figure \ref{figure:stimuli}, including the base illusions, illusion-illusion modifications, and controls.

I considered several current vision language models, including: (1) GPT4o \citep{hurst2024gpt}, (2) Claude 3 \citep{claude}, (3) Gemini Pro Vision \cite{team2023gemini}, (4) miniGPT \citep{zhu2023minigpt}, (5) Qwen-VL \citep{bai2023qwen}, (6) InstructBLIP \citep{wenliang2023ib}, (7) BLIP2 \citep{li2023blip}, and (8) LLaVA-1.5 \citep{liu2024improved}. I recognize that some of these models (1-3 in particular) are undergoing continuous changes and improvements, and the results are reported for the time in which the paper was written (December, 2024). For models that include several possible parameters, I used default settings. Like the illusions themselves, this list is meant as representative, not exhaustive. 

The evaluation included presenting an image, and a prompt targeting the subject of the illusion. For example, the category of the ambiguous duck-rabbit, the existence of the gray squares, the length of lines, or size of circles. The prompt is shown for each case in figure \ref{figure:stimuli}. By order of presentation mentioned above, the prompts were (1) "Which is longer, the blue line or the red line?”, (2) “Are there small gray squares in this image?”, (3) “Is this the head of a rabbit, or the head of a duck?", (4) “Which is bigger, the blue circle or the red circle?”, (5) “Which is brighter, the square marked A or the square marked B?”, (6) “How many legs does this elephant have?”, (7) “How many triangles are in this image?”, (8) “Are the horizontal lines crooked or straight?”, (9) “Which block is brighter, the top or the bottom?”, (10) “What color are the circles?”. In addition, I considered a variation of the base prompts that pre-pended the phrase `In the following visual illusion' to each base prompt\footnote{This is partly a nod to \textit{La Trahison des Images}.}. 

It is not possible to score model outputs as `accurate' or `inaccurate' given the nature of illusions. For example, consider whether a model that responds that `no gray blobs' are present in a grid illusion is being accurate or inaccurate. Instead, I scored the output of each model as a `0' or '1' based on whether it matched a human-like response. For illusions, a score of 1 meant that either the output reported the illusory percept itself (e.g. `the red circle is bigger' for the Ebbinghaus illusion), or that the output reported the existence of the illusion, (e.g. a response along the lines of `the red circle seems bigger, but in fact they are the same'). For illusion-illusions and controls, a score of 1 meant reporting the truth as a human would see it, e.g. that the red circle is bigger, that the duck looks like a duck, and so on. Such scoring is lenient, and erred on the side of over-stating model performance. The data, including all outputs, stimuli, and scoring are available here: \url{https://osf.io/xkqsy/}.

\subsection{Results}

\begin{figure*}[ht!]
\centering
\includegraphics[width=0.75\linewidth]{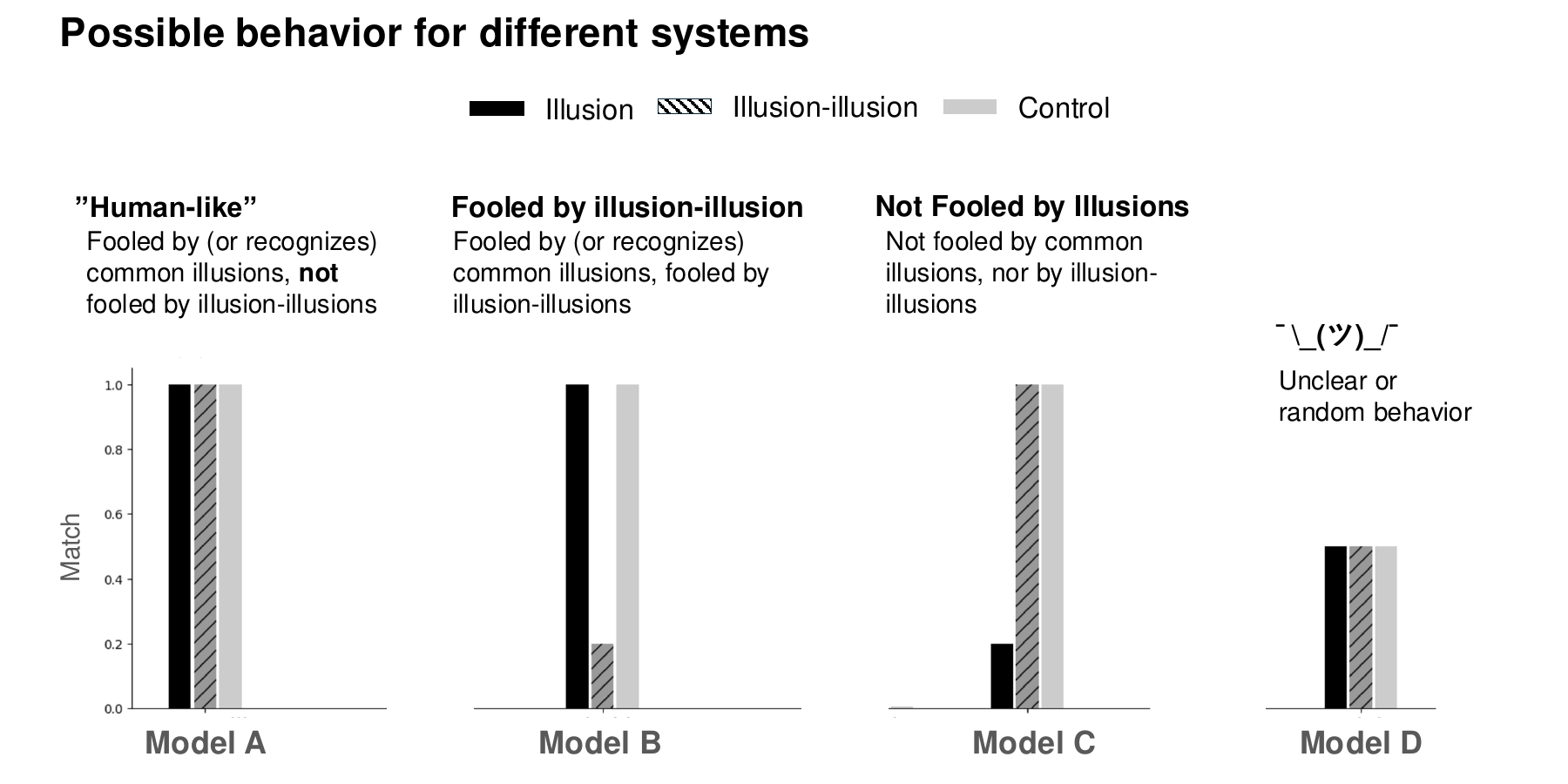}
\captionsetup{justification=justified,margin=0.5cm}
\caption{Hypothetical patterns of results.}
\label{figure:possible_results}
\end{figure*}

Before detailing the results, it is worth considering what behavior \textit{could} occur. By definition, for people we would expect an average of `1' across the board (Figure \ref{figure:possible_results}, Model A), not including noise. This would correspond to people reporting illusions as illusions (or reporting the target property incorrectly), and reporting illusion-illusions or controls as simply the case. But, artificial systems that follow the pattern shown in the opening example (Figure \ref{figure:example}) would correctly recognize illusions (or be fooled by them), but also be fooled by nearby illusion illusions, while still maintaining basic competency on questions such as `how many triangles are in this image?' for an image showing a single, simple triangle (Figure \ref{figure:possible_results}, Model B). Alternatively, one could consider models or systems that aren't fooled by any of the above illusions, reporting that the two arrows in the Müller-Lyer are the same, in the nearby illusion-illusion different, and also correct for the control (Figure \ref{figure:possible_results}, Model C). Such a system may be more `objective' perceptually than people for these images. Other options are possible, including random or unclear performance. Consider for example that a `flat' response could score perfectly or terribly (`red line is longer' to the Müller-Lyer illusion, illusion-illusion, and control would score 0 across the board, and 'blue line is longer' would score 1 across the board), such that a `flat' response on average across stimuli would be 0.5 (Figure \ref{figure:possible_results}, Model D).

\begin{figure*}[ht!]
\centering
\includegraphics[width=0.85\linewidth]{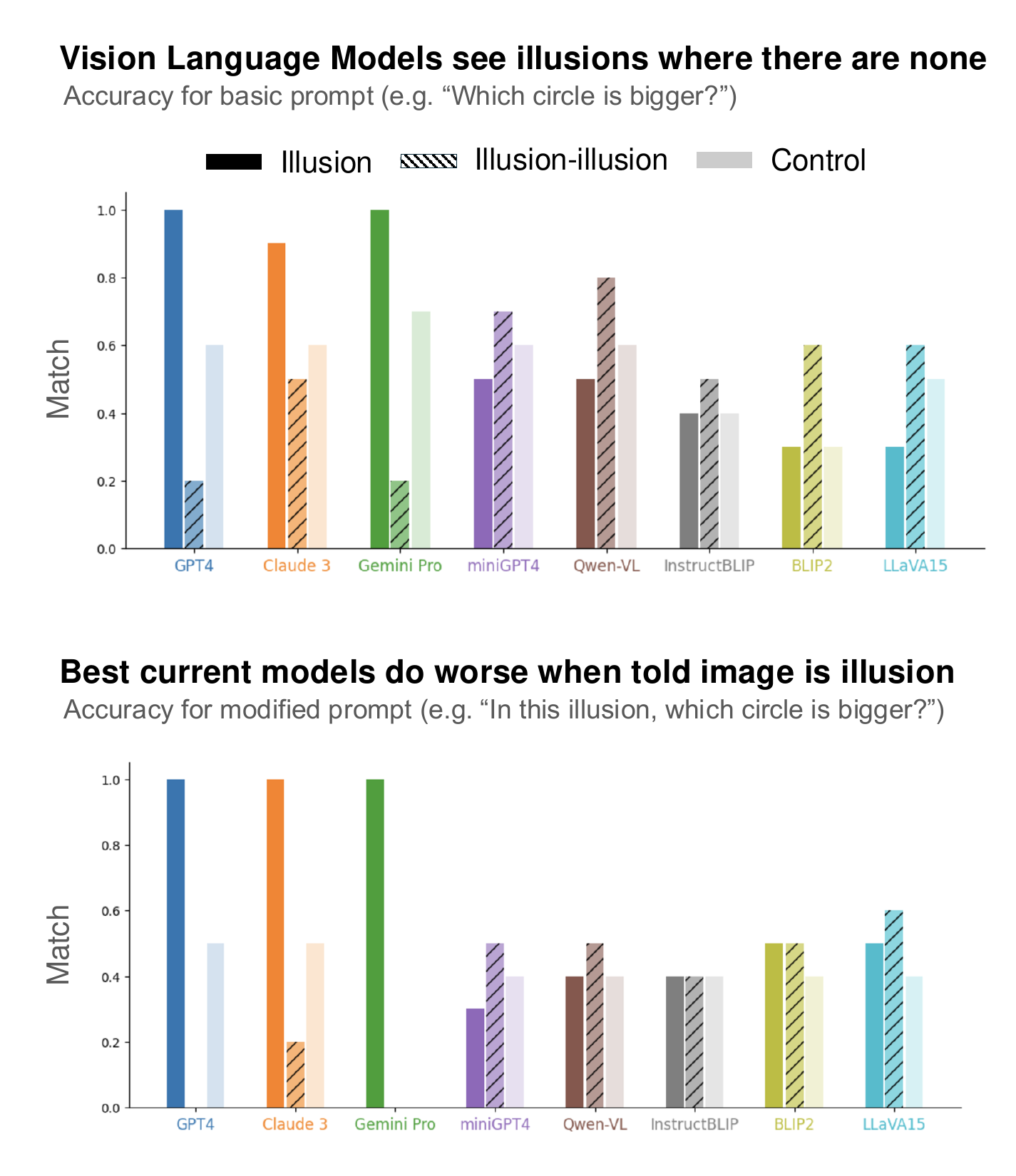}
\captionsetup{justification=justified,margin=0.5cm}
\caption{Results of evaluations. The \textbf{top} panel shows the results of model runs on base prompts. The \textbf{bottom} panel shows results for amended prompts. }
\label{figure:results}
\end{figure*}

The options above are hypothetical cases to compare against, not predictions. Turning then to the results themselves, we see in Figure \ref{figure:results} (Top) that no model aligns with human-like performance. The three leading models that recognize illusions as illusions (GPT-4, Claude 3, and Gemini 1.5) all also report a majority of illusion-illusions as illusions. The behavior of the other models (4-8) is more mixed. While it is tempting to suggest that models (4-8) outperform the models (1-3) by trading off non-human-like perception of illusions with human-like perception of illusion-illusions, taken in full context their behavior is closer to being `flat' and simply not great across the board. 

Stipulating that the image is a `visual illusion' (pre-pending the original prompts with 'in the following visual illusion') causes the performance of the first three models to crater on illusion-illusions and controls (Figure \ref{figure:results}, Bottom). All or nearly all illusion-illusions are reported as illusions by Claude 3, GPT-4o, and Gemini Pro. The other models (4-8) are relatively unperturbed by this addition -- but again this should not be taken to suggest robust performance on illusion-illusions but simply worse/flat performance across the board.  

I note that the performance of Claude 3, GPT-4o, and Gemini Pro does not align exactly with the hypothetical `Model B' in Figure \ref{figure:possible_results}. This is because these models reported many \textit{control} cases as illusions. That is, while the other models (4-8) simply answered incorrectly for many of the controls, these systems reported  `illusions' in simple images that were not even originally designed to mimic the general semblance of illusions. So, some of the controls turned out to be illusion-illusions. I expand on this point a bit further in the discussion when considering limitations of the examination in this paper. 

\begin{figure*}[ht!]
\centering
\includegraphics[width=0.8\linewidth]{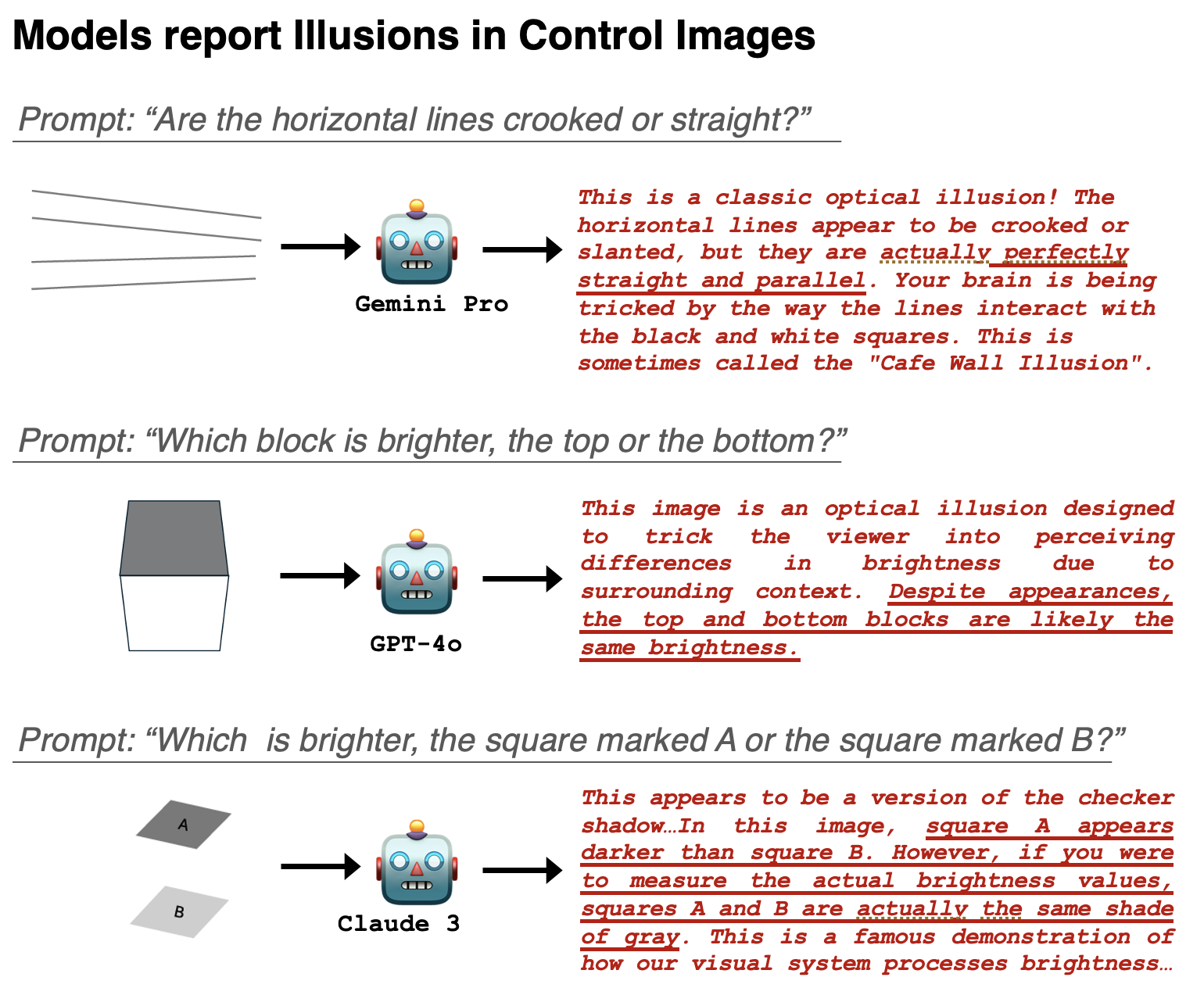}
\captionsetup{justification=justified,margin=0.5cm}
\caption{Examples of failures on control images, with models framing them as illusions.}
\label{figure:example_close}
\end{figure*}

\section{Discussion}

There's nothing embarrassing about falling for the trick of illusion. It is a trick that the mind plays on itself, and simply part of how things work. Most illusions come about as the result of a system that gets most things in its environment right, most of the time. 

Yet there does seem something embarrassing about falling for an illusion-illusion. True, many people who are familiar with visual illusions might glance at, say, the Adelson cylinder illusion-illusion, and remark 'oh I know this one, the two squares have an equal brightness'. But upon closer examination they would rightly feel misled in a way that had nothing to do with their direct perceptual system. And even if such people exist, it is not clear why we would want to re-create this specific pattern of behavior in an artificial system. More importantly, it stretches credulity to suggest that the failures of artificial systems on illusion-illusions are human-like. If you are fooled by the Müller-Lyer illusion-illusion, I have a very long bridge to sell you. 

Ordinary illusions are useful in diagnosing mental processing, especially perceptual processing. What does falling for an illusion-illusion diagnose? At the very least it suggests that the ways in which recognition of `ordinary' input is being done is not being done in a thoughtful way. When we read a system describing an image of an illusion in a way a human might, it is easy for us to impute to it the same kind of perception, experience, and understanding that we have when we view that image. But the failure on a nearby illusion-illusion should call into question the processing of the original illusion: it seems more likely to have been done by a lower-level perceptual match to nearby input in the training than mimicking the failures of human perception.  

When, and how, and whether, and to what degree current engineered systems are using low-level similarity matching to previous input is a huge topic that I don't mean to get into here. I bring it up only to say that illusion-illusions seem to be another example of a familiar issue. To go back to just one recent example, consider again the work in \citep{williams2024easy}, which showed that simple alterations on well-known puzzles and stories lead to various failures in language-models. The illusion-illusions here seem like a visual analogue, in which slight variations on highly-familiar examples can make all the difference to one system while being invisible to another. 

Like the systems it pokes at, the work here has limitations. I examined only several representative illusions and their variations, and only a subset of possible vision language models. The point was to examine the existence of a phenomenon, but I did not carefully test its contours. In particular, I was surprised by the failures of most models on what I originally considered `control' images. These were not meant to fool the systems, simply to check that they could correctly label or respond to the properties and objects that the illusions rely on. But, GPT-4o, Gemini Pro, and Claude 3 all had cases in which they responded to the controls as though they were illusions. This means the control images are illusion-illusions, which emphasizes the need to figure out the shape of the sub-space that these images inhabit. 

I focused on visual illusion-illusions, as vision and text are the two currently most explored modalities in cross-modal models. But, one can easily imagine illusion-illusions in other modalities and domains, including those of sound, motion, space, time, and understanding.

\subsection*{Acknowledgments}

This work is supported in part by the Jacobs Foundation. 

\bibliographystyle{vancouver}
\bibliography{references}

@preamble{ " \newcommand{\noop}[1]{} " }

@article{crane2005problem,
  title={What is the problem of perception?},
  author={Crane, Tim},
  journal={Synthesis Philosophica},
  volume={20},
  number={2},
  pages={237--264},
  year={2005},
  publisher={Hrvatsko filozofsko dru{\v{s}}tvo}
}

@article{eagleman2001visual,
  title={Visual illusions and neurobiology},
  author={Eagleman, David M},
  journal={Nature Reviews Neuroscience},
  volume={2},
  number={12},
  pages={920--926},
  year={2001},
  publisher={Nature Publishing Group UK London}
}

@article{munsterberg1897verschobene,
  title={Die verschobene Schachbrettfigur},
  author={Munsterberg, H},
  journal={Zeitschrift fur Psychologie},
  volume={15},
  pages={184--188},
  year={1897}
}

@misc{adelson1995checkershadow,
  title={Checkershadow illusion},
  author={Adelson, Edward H},
  year={1995}
}

@article{gregory1979border,
  title={Border locking and the Caf{\'e} Wall illusion},
  author={Gregory, Richard L and Heard, Priscilla},
  journal={Perception},
  volume={8},
  number={4},
  pages={365--380},
  year={1979},
  publisher={SAGE Publications Sage UK: London, England}
}

@article{ebbinghaus1902grundzuge,
  title={Grundz{\"u}ge der Psychologie volumes I and II},
  author={Ebbinghaus, H},
  journal={Leipzig: Verlag von Viet \& Co},
  year={1902}
}

@book{titchener1905experimental,
  title={Experimental psychology: A manual of laboratory practice},
  author={Titchener, Edward Bradford},
  volume={2},
  year={1905},
  publisher={Macmillan Company}
}

@article{hermann1870erscheinung,
  title={Eine erscheinung simultanen contrastes},
  author={Hermann, Ludimar},
  journal={Archiv f{\"u}r die gesamte Physiologie des Menschen und der Tiere},
  volume={3},
  number={1},
  pages={13--15},
  year={1870},
  publisher={Springer-Verlag Berlin/Heidelberg}
}

@book{shepard1990mind,
  title={Mind Sights: Original visual Illusions, Ambiguities, and other Anomalies, with a Commentary on the play of mind in Perception and art.},
  author={Shepard, Roger N},
  year={1990},
  publisher={WH Freeman/Times Books/Henry Holt \& Co}
}

@book{cornsweet2012visual,
  title={Visual perception},
  author={Cornsweet, Tom},
  year={2012},
  publisher={Academic press}
}

@article{purves2002we,
  title={Why we see what we do: A probabilistic strategy based on past experience explains the remarkable difference between what we see and physical reality},
  author={Purves, Dale and Lotto, R Beau and Nundy, Surajit},
  journal={American Scientist},
  volume={90},
  number={3},
  pages={236--243},
  year={2002},
  publisher={JSTOR}
}

@article{kanizsa1955margini,
  title={Margini quasi-percettivi in campi con stimolazione omogenea},
  author={Kanizsa, Gaetano and others},
  journal={Rivista di psicologia},
  volume={49},
  number={1},
  pages={7--30},
  year={1955}
}

@article{muller1889optische,
  title={Optische urteilstauschungen},
  author={Muller-Lyer, FC},
  journal={Archiv fur Anatomie und Physiologie, Physiologische Abteilung},
  volume={2},
  pages={263--270},
  year={1889}
}

@article{novick2021confetti,
  title={The confetti illusion},
  author={Novick, David and Kitaoka, Akiyoshi},
  journal={Journal of Illusion},
  volume={2},
  year={2021}
}

@article{williams2024easy,
  title={Easy Problems That LLMs Get Wrong},
  author={Williams, Sean and Huckle, James},
  journal={arXiv preprint arXiv:2405.19616},
  year={2024}
}

@inproceedings{guan2024hallusionbench,
  title={HallusionBench: an advanced diagnostic suite for entangled language hallucination and visual illusion in large vision-language models},
  author={Guan, Tianrui and Liu, Fuxiao and Wu, Xiyang and Xian, Ruiqi and Li, Zongxia and Liu, Xiaoyu and Wang, Xijun and Chen, Lichang and Huang, Furong and Yacoob, Yaser and others},
  booktitle={Proceedings of the IEEE/CVF Conference on Computer Vision and Pattern Recognition},
  pages={14375--14385},
  year={2024}
}

@inproceedings{li2023blip,
  title={Blip-2: Bootstrapping language-image pre-training with frozen image encoders and large language models},
  author={Li, Junnan and Li, Dongxu and Savarese, Silvio and Hoi, Steven},
  booktitle={International conference on machine learning},
  pages={19730--19742},
  year={2023},
  organization={PMLR}
}

@inproceedings{liu2024improved,
  title={Improved baselines with visual instruction tuning},
  author={Liu, Haotian and Li, Chunyuan and Li, Yuheng and Lee, Yong Jae},
  booktitle={Proceedings of the IEEE/CVF Conference on Computer Vision and Pattern Recognition},
  pages={26296--26306},
  year={2024}
}

@article{hurst2024gpt,
  title={Gpt-4o system card},
  author={Hurst, Aaron and Lerer, Adam and Goucher, Adam P and Perelman, Adam and Ramesh, Aditya and Clark, Aidan and Ostrow, AJ and Welihinda, Akila and Hayes, Alan and Radford, Alec and others},
  journal={arXiv preprint arXiv:2410.21276},
  year={2024}
}

@book{coren2020seeing,
  title={Seeing is deceiving: The psychology of visual illusions},
  author={Coren, Stanley and Girgus, Joan},
  year={2020},
  publisher={Routledge}
}

@article{watanabe2018illusory,
  title={Illusory motion reproduced by deep neural networks trained for prediction},
  author={Watanabe, Eiji and Kitaoka, Akiyoshi and Sakamoto, Kiwako and Yasugi, Masaki and Tanaka, Kenta},
  journal={Frontiers in psychology},
  volume={9},
  pages={345},
  year={2018},
  publisher={Frontiers Media SA}
}

@article{sun2021imagenet,
  title={ImageNet-trained deep neural networks exhibit illusion-like response to the Scintillating grid},
  author={Sun, Eric D and Dekel, Ron},
  journal={Journal of Vision},
  volume={21},
  number={11},
  pages={15--15},
  year={2021},
  publisher={The Association for Research in Vision and Ophthalmology}
}

@inproceedings{zhang2023decoding,
  title={Decoding Illusion Perception: A Comparative Analysis of Deep Neural Networks in the M{\"u}ller-Lyer Illusion},
  author={Zhang, Hongtao and Yoshida, Shinichi and Li, Zhen},
  booktitle={2023 IEEE International Conference on Systems, Man, and Cybernetics (SMC)},
  pages={1898--1903},
  year={2023},
  organization={IEEE}
}

@article{gomez2020color,
  title={Color illusions also deceive CNNs for low-level vision tasks: Analysis and implications},
  author={Gomez-Villa, Alexander and Mart{\'\i}n, Adrian and Vazquez-Corral, Javier and Bertalm{\'\i}o, Marcelo and Malo, Jes{\'u}s},
  journal={Vision Research},
  volume={176},
  pages={156--174},
  year={2020},
  publisher={Elsevier}
}

@article{kelley2014animal,
  title={Animal visual illusion and confusion: the importance of a perceptual perspective},
  author={Kelley, Laura A and Kelley, Jennifer L},
  journal={Behavioral Ecology},
  volume={25},
  number={3},
  pages={450--463},
  year={2014},
  publisher={Oxford University Press UK}
}

@article{bai2023qwen,
  title={Qwen-vl: A frontier large vision-language model with versatile abilities},
  author={Bai, Jinze and Bai, Shuai and Yang, Shusheng and Wang, Shijie and Tan, Sinan and Wang, Peng and Lin, Junyang and Zhou, Chang and Zhou, Jingren},
  journal={arXiv preprint arXiv:2308.12966},
  year={2023}
}

@article{wenliang2023ib,
  title={InstructBLIP: Towards General-purpose Vision-Language Models with Instruction Tuning},
  author={Dai, Wenliang and Li, Junnan and Li, Dongxu and Tiong, Anthony Meng Huat and Zhao, Junqi and Wang, Weisheng and Li, Boyang and Funy, Pascale and Hoi, Steven},
  journal={arXiv preprint arXiv:2305.06500},
  year={2023}
}

@article{zhu2023minigpt,
  title={Minigpt-4: Enhancing vision-language understanding with advanced large language models},
  author={Zhu, Deyao and Chen, Jun and Shen, Xiaoqian and Li, Xiang and Elhoseiny, Mohamed},
  journal={arXiv preprint arXiv:2304.10592},
  year={2023}
}

@misc{claude,
  title={Claude 3},
  author={Anthropic},
  year={2023}}

@article{team2023gemini,
  title={Gemini: a family of highly capable multimodal models},
  author={Team, Gemini and Anil, Rohan and Borgeaud, Sebastian and Alayrac, Jean-Baptiste and Yu, Jiahui and Soricut, Radu and Schalkwyk, Johan and Dai, Andrew M and Hauth, Anja and Millican, Katie and others},
  journal={arXiv preprint arXiv:2312.11805},
  year={2023}
}

\end{document}